\documentclass[11pt]{article}
\usepackage{moriond,epsfig}

\begin{document}
\vspace*{4cm}
\title{A MINIMAL MODEL WITH LARGE EXTRA DIMENSIONS TO FIT THE NEUTRINO DATA}

\author{LING FU-SIN}

\address{Service de Physique Th\'eorique\\
         Universit\'e Libre de Bruxelles\\
		Campus de la Plaine CP225\\
		Bvd du Triomphe, 1050 Brussels, Belgium}

\maketitle\abstracts{
The existence of large extra dimensions enables to explain very simply the 
various neutrino suppression data. The flavour neutrinos, confined on the brane,
are coupled to a "bulk" neutrino state, which is seen as an infinite tower of 
Kaluza-Klein states from the 4-dimensional point of view. The oscillation of
flavour neutrinos into these states can lead to a global and sizable 
energy-independent suppression at large distances, which can be related 
to the solar neutrino puzzle.~\cite{our}}

\section{Introduction}

The possibility of "large extra dimensions", i.e. with (at least) one
compactification radius close to the current validity limit of Newton's 
law of gravitation ($\sim$ 1mm), raises considerable interest. 
One reason is that it could solve the "so-called" hierarchy problem~\cite{led} 
by providing us with a new fundamental energy scale $M_*$ around 1~TeV. 
The discrepancy between the apparent huge value of the Planck scale and the weak scale 
then results from the fact that extra dimensions remain "hidden" at low energies for 
all interactions except gravitation. From the relation
\[
M^2_{Planck}=V_n \cdot M^{n+2}_*
\]
we see that at least 2 extra dimensions are needed ($n \geq 2$) to avoid deviations 
from Newton's law.  

Another reason is that it can bring some new physics just "around the corner", at energies 
as low as 1 TeV. In particular, neutrino physics is a ideal area to study 
new theories. While particles belonging to the Standard Model (SM) are required to be 
confined on the usual four dimensional spacetime 
(the 3-brane with thickness~$\sim 1~TeV^{-1}$), sterile neutrinos, 
which do not experience any of the gauge interactions, can propagate in the 4+n 
dimensional spacetime as well as gravitons, and turn to be an efficient 
tool to probe the "bulk" of space. 

From the four dimensional point of view, bulk neutrinos appear as an infinite tower 
of states, called the Kaluza-Klein states~\cite{lednu}, as a result of the compactification of the extra
dimensions. When coupled to SM flavour neutrinos, unconventional patterns of neutrino masses
and mixings arise.
Recent works~\cite{dvali,ramond,lorenzana,creminelli} have shown that  
it is, at least partially, possible to accommodate experimental constraints 
on neutrinos within this setup.

A common feature of these solutions is that the solar $\nu _e$ depletion is explained
using a small mixing between flavour and bulk neutrinos, which is enhanced in the sun
by matter effects, the so-called Mikheyev-Smirnov-Wolfenstein (MSW)mechanism. 
These solutions are analogous to the classical Small Mixing Angle oscillation solution 
(SMA), and are strongly energy-dependent.
However, recent analysis of SuperKamiokande (SK) data disfavours this kind of solutions, as
no spectral distortion, nor seasonal effect, nor day-night effect are observed. Therefore, 
a global 40-60\% suppression~\footnote{However, the Homestake experiment observes 
even more neutrino suppression, the observed to expected flux ratio is around 33\%}, 
which is energy and distance independent, is perfectly 
compatible with the observed deficit.

Using the setup of large extra dimensions, we explore this possibility in details.
We also require that no MSW effects are present, otherwise, inescapable resonances 
in the sun could kill the final large distance survival probability. Finally, we try
to accommodate other experimental constraints on neutrinos, including
the depletion of atmospheric $\nu _\mu$, the $\overline{\nu }_{e}$ no disappearance at Chooz,
the $\nu _\mu$ disappearance at K2K, $\overline{\nu }_{e}$ no appearance at Karmen 
and the LSND signal~\footnote{For details on the discussion of the 
experimental constraints, see Ref.~\cite{our} and references therein}.

\section{The 1-flavour model.}

As a first attempt, we will investigate a toy model in which only 1 flavour neutrino 
$\nu _1$ (either of $\nu _e$, $\nu _\mu $, $\nu _\tau $) 
is coupled to one massless bulk fermion $\Psi$ through a Yukawa coupling constant $\lambda$.
Moreover, we consider that only one extra dimension, taken as a circle of radius $R$, 
is relevant from the four dimensional point of view 
({\it i.e.} all other extra dimensions are assumed to be much smaller).

The 5D action used is the following~\cite{dvali,creminelli} : 
\begin{equation}
\label{action}
S=\int d^4x\,dy\;\overline{\Psi} i\Gamma _A\partial ^A\Psi +\int d^4x\{\overline{\nu} _1 
i\gamma _\mu \partial ^\mu \nu _1+\lambda \overline{\nu} _1\Psi (x^\mu ,y=0)H(x^\mu
)+h.c.\} 
\end{equation}
where $A=0,...,4$ and $x^4=y$ is the extra dimension. 
The Yukawa coupling between the usual Higgs scalar $H$, 
the weak eigenstate neutrino $\nu _1$ and the bulk
fermion  only operates at $y=0$, {\it i.e.} on the 3-brane.

Expanding the bulk fermion into Fourier modes (or Kaluza-Klein states),
\begin{eqnarray*}
\Psi (x^\mu,y)=\sum _{n=-\infty}^{+\infty}\frac 1{\sqrt{2\pi R}}\psi _n (x^\mu)e^{i n y/R}
\end{eqnarray*}
and replacing $H$ by its vacuum expectation value $v$,
one ends up with an infinite mass matrix with eigenvalues $\frac{\lambda _n}R$ given
by the characteristic equation $\lambda _n =\pi \xi ^2\cot (\pi \lambda _n)$, 
where $\xi \equiv m R$ and $m = \frac{\lambda v}{\sqrt{2 \pi R}}$.
Mixings $U_{0n}$ between the flavour neutrino and the mass eigenstates 
$\nu _{\lambda _n}$ are given by 
\begin{eqnarray}
\left| \nu _1 \right\rangle =\sum _{n=0}^{\infty }U_{0n}
\left| \nu _{\lambda _n} \right\rangle 
\hspace{.5in}{\rm and} \hspace{.5in}
U_{0n}^2 =\frac 2{1+\pi^2 \xi ^2+ \frac{\lambda _n ^2}{\xi ^2}}
\end{eqnarray}

The survival amplitude $A_{\nu _1\nu _1}$and the survival probability 
$P_{\nu _1\nu _1}$ are given by 

\begin{eqnarray}
\label{anu1nu1}
A_{\nu _1\nu _1} &=&\sum _{n=0}^{\infty }\left( U_{0n}\right)
^2e^{\,i\left( \,\lambda _n\right) ^2x} \\
\label{pnu1nu1}
P_{\nu _1\nu _1} &=&\sum _{n=0}^{\infty }\left( U_{0n}\right) ^4+\sum
\sum _{n \neq m}\left( U_{0n}\right) ^2\left( U_{0m}\right) ^2\cos
\left[ \left( \left( \,\lambda _n\right) ^2-\left( \,\lambda _m\right)
^2\right) x\right]
\end{eqnarray}
where $x = \frac{L}{2ER^2} \approx 10^{-7}
\frac{\left( L/km\right) }{(E/GeV)(R/mm)^2}$.

From  Eq.~(\ref{pnu1nu1}), it is straightforward that the mean value~\footnote{The mean value is understood 
here as the average over a large interval in $x$.} of  the survival probability and the amplitude of its
fluctuations are given by 

\begin{eqnarray}
\label{meanprob}
\left< P_{\nu _1 \nu _1} \right> &=& \sum _{n=0}^{\infty} \left( U_{0n}\right) ^4 \\
\label{fluctu}
\sigma ^2 \left( P \right) &=& \left( \sum _{n=0}^{\infty} \left( U_{0n}\right) ^4 \right) ^2 
- \sum _{n=0}^{\infty} \left( U_{0n}\right) ^8
\end{eqnarray}
The mean value $\left< P_{\nu _1 \nu _1} \right>$ is dominated by the zero-mode contribution $\left(
U_{00}\right) ^4$ for $\xi \leq 1/3$, while the large $\xi$ regime, 
$\left< P_{\nu _1 \nu _1} \right> = \frac{1}{\pi^2 \xi^2}$
is entered from $\xi \approx 0.8$. 
At large $\xi$, the amplitude of the fluctuations $\sigma \left( P \right)$ 
tends asymptotically to $\left< P_{\nu _1 \nu _1} \right>$ (see Fig.~1).

It is worth noting that the eigenvalues $\lambda _n$ are roots of a transcendental equation.
Although they are close to integers, thus quasi-harmonic, a periodic behaviour for the 
survival probability is not recovered, especially at large $\xi$. This is also clear 
form the values of $\left< P_{\nu _1 \nu _1} \right>$ and 
$\sigma \left( P \right)$, as can be seen in Fig.~1.
Once a flavour neutrino starts to "oscillate" with Kaluza-Klein states, it never 
reappears as a pure flavour state. 
 
\begin{figure}
\centerline{\epsfig{figure=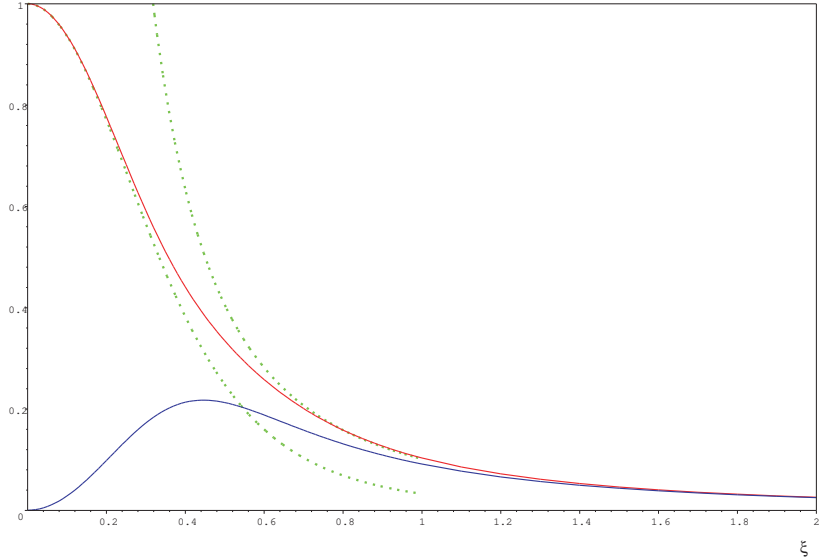,height=3in}}
\caption{\label{meanprob}
Mean survival probability $\left< P _{\nu _1 \nu _1} \right>$ 
and fluctuations $\sigma(P)$ as functions of $\xi$. 
The dashed lines give the small $\xi$ ($\left< P _{\nu _1 \nu _1} \right> \approx (U_{00})^4)$,
and the large $\xi$ ($\left< P _{\nu _1 \nu _1} \right> \approx \frac{1}{\pi^2 \xi^2}$) 
approximations.}
\end{figure}

The simplest toy model can now be compared to experimental data.
For $\nu _e$, we are looking for a global 40 to 60\% suppression at large $x$, 
as the simplest interpretation of the absence of $L/E$ dependence in 
the SuperKamiokande solar neutrinos data. 
This is possible because the Sun-Earth system is a very long-baseline system.
As the solar core is large (typically, $x \sim 10^5$ and $\Delta x \sim 10^2 >> 1$ 
for solar neutrinos with $E \sim 1 MeV$ and $R \sim 1 mm$), 
fluctuations are completely washed away and  
the only observable effect will be an average suppression.
Values of $\xi$ with $0.29< \xi < 0.42$ fit the solar $\nu _e$ data.

However, we still should avoid MSW effects in the Sun or Earth.
It turns out that this requirement and the Chooz constraint cannot be 
simultaneously respected. 

The Chooz nuclear reactor experiment shows no $\overline{\nu }_{e}$ disappearance 
at a distance $L=1~km$ and a typical energy of 2 MeV. 
At fixed $\xi$, we can extract the largest allowed $x$ value for Chooz 
still fitting the data, or equivalently an upper limit for $1/R^2$.
As $1/R$ controls the typical mass difference between two
consecutive Kaluza-Klein levels, MSW resonant conversion will
take place if $1/R$ is small compared to the MSW potential $V_{MSW}$. 
More precisely, this will happen if $1/R^2 \leq 2 E V_{MSW}$.
Conversely, MSW effects can be avoided by putting a lower limit to $1/R^2$,
which can then turn to be incompatible with the previous upper limit deduced from
the Chooz constraint.
The Fig.~2 shows that this is indeed the case in the toy model.

\begin{figure}
\centerline{\epsfig{figure=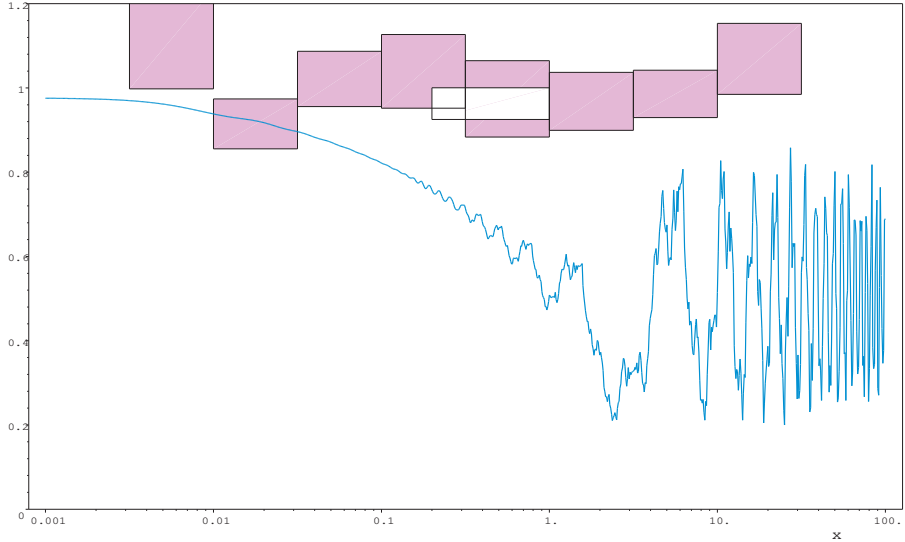,height=3in}}
\caption{\label{nue11}
Comparison of the toy model with the $\nu _e$ constraints.
The mean survival probability at large $x$ is fixed at 50\% to fit the solar neutrinos.
We have plotted the latest available experimental data we could find.
The series of filled boxes show the $L/E$ dependence for the atmospheric $\nu _e$, 
as observed by SK (1$\sigma$).
The data have been normalized by an overall 
0.95 factor with respect to the raw data 
(instead of the usual $\sim$ 0.9 used by SK).
The open box corresponds to the Chooz constraint.  
The error bar at $2\sigma$ level combines quadratically the statistical and systematic 
errors. The boxes cannot slide to the
left without getting into the range of the MSW effect, showing thus a complete
disagreement with the toy model.}
\end{figure}

\begin{figure}
\centerline{\epsfig{figure=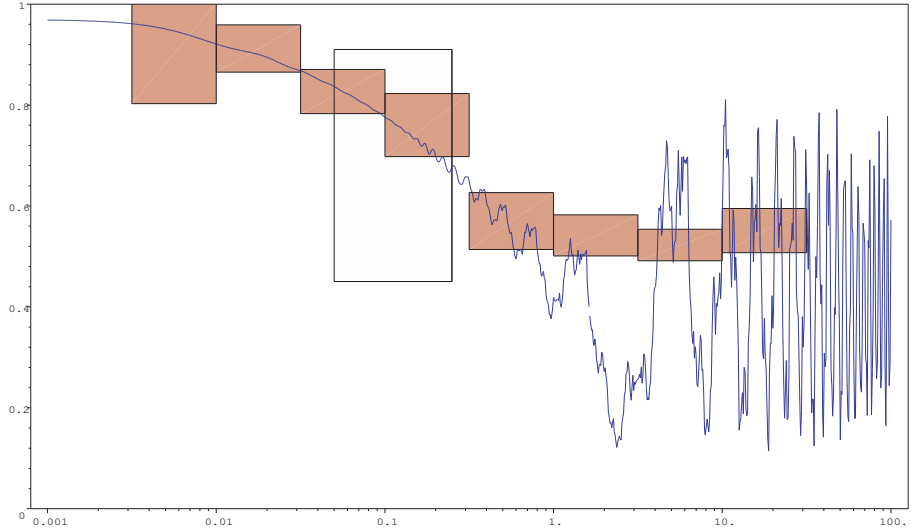,height=3in}}
\caption{\label{numu11}
Comparison of the toy model with the $\nu _\mu$ constraints ($\xi=4/10$).
The series of filled boxes show the $L/E$ dependence for the atmospheric $\nu _\mu$, 
as observed by Superkamiokande (1$\sigma$; the same 0.95 factor has been used). 
The open box corresponds to the K2K constraint. The error bar at $1\sigma$ level
combines quadratically the statistical and systematic errors. 
The survival probability is in good agreement with
the constraints, once an average on each energy 'bin' is performed.}
\end{figure}

One can also try to fit the $\nu _{\mu }$ data in the toy model (that is $\nu _1=\nu _\mu$). 
The Fig.~\ref{numu11} shows that a good agreement with the experimental data can be obtained 
for $\xi=0.4$.

\section{The 2-flavour model.}

While almost as simple as the previous toy model, the 2-flavour model
enables to accommodate most of the neutrino experimental data. 

In addition to $\nu _e$, a second flavour neutrino, taken to be $\nu _\mu$,
is included in the model. Only one linear combination
of them, namely $\cos \theta \;\nu _{e}+\sin \theta \;\nu _{\mu }$
($\theta$ is a new parameter : the mixing angle), 
which we call as previously $\nu _{1}$, 
is coupled to the bulk fermion. 
The Yukawa coupling proceeds exactly as in Eq.~\ref{action}.
The orthogonal linear combination,
$\nu _{2} =-\sin \theta \;\nu _{e}+\cos \theta \;\nu _{\mu }$ remains massless 
and decouples.

The phenomenology of the 2-flavour model can be quite different from that of 
the toy model. Indeed, the mixing angle plays a crucial role in the survival
probabilities of the flavour neutrinos. 
\begin{eqnarray*}
P_{\nu _{e}\nu _{e}} &=&\cos ^{4}\theta \;P_{\nu _{1}\nu _{1}}+\sin
^{4}\theta +2\sin ^{2}\theta \;\cos ^{2}\theta \;\textrm{Re}\left( A_{\nu
_{1}\nu _{1}}\right) \\
P_{\nu _{\mu }\nu _{\mu }} &=&\sin ^{4}\theta \;P_{\nu _{1}\nu _{1}}+\cos
^{4}\theta +2\sin ^{2}\theta \;\cos ^{2}\theta \;\textrm{Re}\left( A_{\nu
_{1}\nu _{1}}\right)
\end{eqnarray*}
Moreover, a transition $\nu _e \leftrightarrow \nu _\mu$ becomes possible and its 
probability is given by
\[
P_{\nu _{e}\nu _{\mu }}=P_{\nu _{\mu }\nu _{e}}=\sin ^{2}\theta \ \cos
^{2}\theta \ (P_{\nu _{1}\nu _{1}}-2\textrm{Re}\left( A_{\nu _{1}\nu
_{1}}\right) +1) 
\]
The 2-flavour model has 3 degrees of freedom, $(\xi ,\theta ,R)$, 
to fit the experimental data. 

\begin{figure}
\centerline{\epsfig{figure=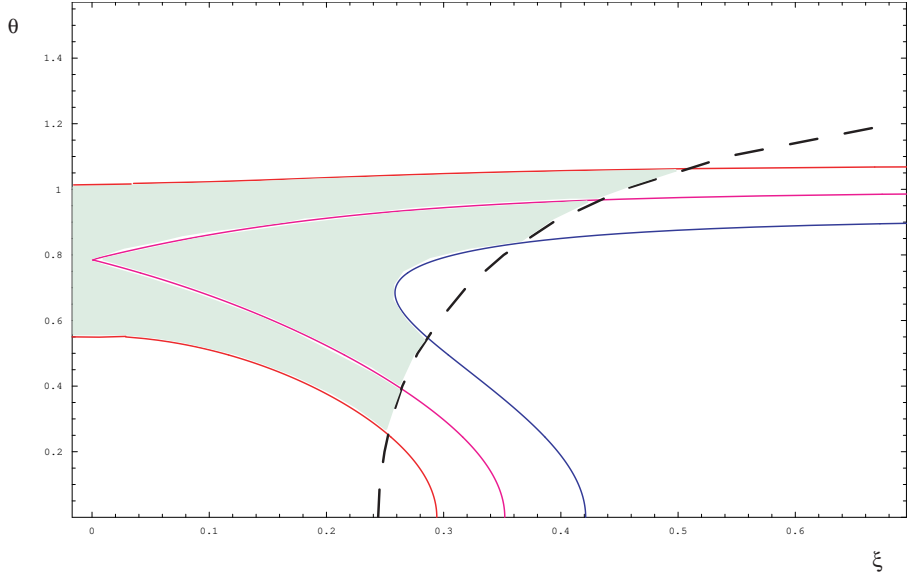,height=3in}}
\caption{\label{theta}
Region of values for $\xi$ and $\theta$ for which the solar 
and Chooz constraints for $\nu _e$ can be accommodated without MSW effect.
The solid lines correspond (from top to bottom) to a solar neutrino mean survival 
probability of 60, 50, and 40\%. The dashed line 
indicates the frontier of the MSW region in view of the Chooz constraint. 
The filled region gives the allowed $(\xi,\theta)$ values.}
\end{figure}

We first focus on the electronic neutrino data. 
As before, a global 40 to 60\% suppression can account for 
the solar neutrino deficit.
The contour plot of the mean $\nu _e$ survival probability 
in the plane $\xi -\theta$ is shown in Fig.~\ref{theta}, 
and defines a region of allowed values for the a-dimensional 
coupling $\xi$ and the mixing angle $\theta$. 

Again, the Chooz constraint together with the no-MSW requirement forbids
certain values for $(\xi,\theta)$, as can be seen in Fig.~\ref{theta}. A large allowed 
region of the parameter space still enables to fit all the $\nu _e$ experimental 
constraints. All solutions have a rather large mixing angle $\theta \sim \pi/4$.
We can also verify that a mixing angle $\theta=0$ is indeed excluded. 

\begin{figure}
\centerline{\epsfig{figure=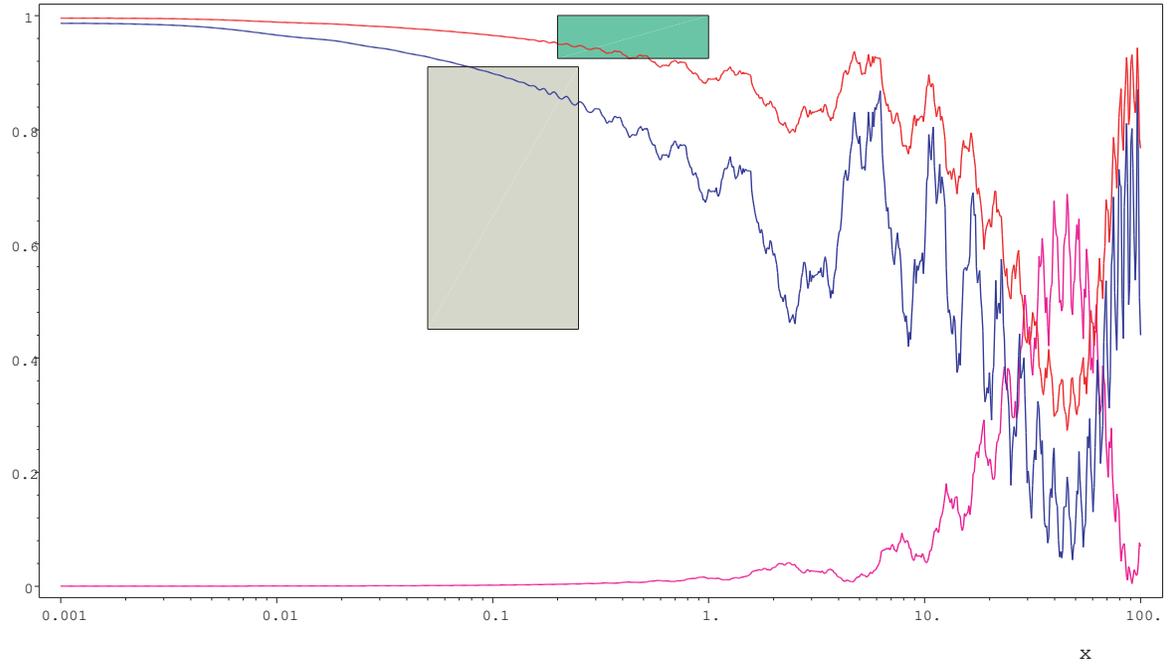,height=3.5in}}
\caption{\label{nu21}
Comparison of the 2-flavour model with the Chooz and K2K constraints.
(highest curve [red] is for $\nu _e$)
$\xi=0.3$ and $\theta=1.05$, so that $\left<P_{\nu _e \nu _e} \right> \simeq 60\%$. 
The transition probability $P_{\nu _e \nu _\mu}$ is also depicted (lowest pink curve).}
\end{figure}

We can now discuss the constraints involving the muonic neutrinos.
The $\nu _\mu $ disappearance experiment K2K, reveals some 30\% deficit 
for 2 GeV neutrinos at a distance $L\simeq 250\ km$. 
As $x_{K2K}\simeq 1/4 \cdot x_{Chooz}$, muonic neutrinos are expected to
disappear more than electronic neutrinos. This requires $\theta >\pi /4$,
and higher values of $\xi $ are preferred, as $P_{\nu _{e}\nu _{e}}-P_{\nu
_{\mu }\nu _{\mu }}\propto (1-P_{\nu _{1}\nu _{1}})$. However, even for the
maximal allowed $\xi $, the preliminary result of K2K can only be accommodated by taking
the large statistical error into account. 
The Fig.~\ref{nu21} shows a possible fit for $\nu _e$ and $\nu _\mu$, 
which solves the solar neutrino problem, and simultaneously
satisfies the Chooz and K2K constraints. 
It will be shown hereafter that this solution also fits the atmospheric neutrinos data. 

\begin{figure}
\centerline{\epsfig{figure=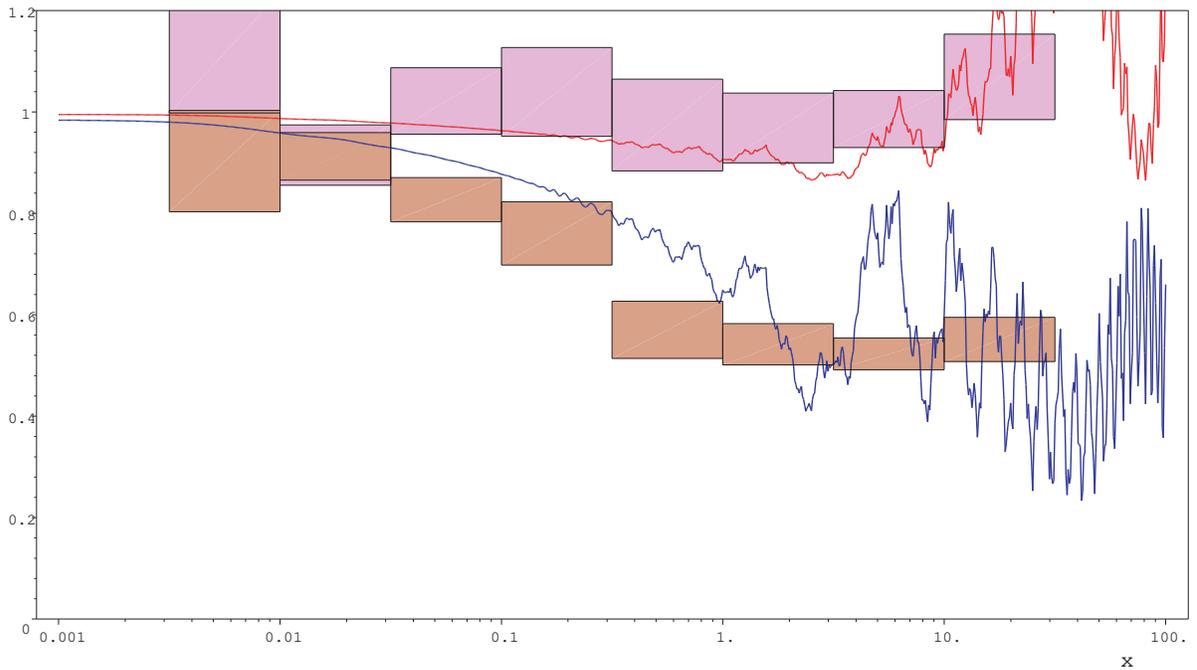,height=3.5in}}
\caption{\label{nuSK21}
Expected atmospheric neutrino fluxes in the 2-flavour model and SK data. 
$\xi=0.3$ and $\theta=1.05$ as in Fig.~\ref{nu21}. The initial flux 
$2 \stackrel {( - )}{\nu _\mu }+ \stackrel {( - )}{\nu _e }$ is 
normalized to 1 at $x=0$. The observed SK data has been normalized as in Fig.~\ref{nue11}
The agreement with experimental data is quite remarkable.}
\end{figure}

To discuss the constraints coming from the atmospheric neutrinos, we recall that in the 
2-flavour model, a transition $\nu _e \rightarrow \nu _\mu$ or $\nu _\mu \rightarrow \nu _e$ 
is possible. The transition probability, as shown in Fig.~\ref{nu21}, becomes non negligible
in the range of the atmospheric neutrinos. As the atmospheric neutrinos originate from the 
decay of the charged pions and kaons into muons and the subsequent decay of muons into  
electrons, the ratio of the neutrino initial fluxes 
$\frac{\phi ^{(i)}_{\nu _\mu}}{\phi ^{(i)}_{\nu _e}}$ 
is expected to be very close to 2, especially at low energy~\footnote{at higher
energy, the produced muon can go through the atmosphere without decaying, so that the
ratio $\frac{\phi ^{(i)}_{\nu _\mu}}{\phi ^{(i)}_{\nu _e}}$ increases with energy}. 
Therefore, the expected atmospheric neutrino flux in the 2-flavour model 
is given by (we don't distinguish between $\nu$ and $\overline{\nu}$)
\begin{eqnarray}
\frac{\phi _{\nu _e}}{\phi ^{(i)}_{\nu _e}}=P_{\nu _e \nu _e}+ 2 \, P_{\nu _\mu \nu _e} \\
\frac{\phi _{\nu _\mu}}{\phi ^{(i)}_{\nu _\mu}}=P_{\nu _\mu \nu _\mu}
+ 1/2 \, P_{\nu _e \nu _\mu}
\end{eqnarray}
As a result, the observed $\nu _e$ flux can be enhanced compared to the initial production 
flux. In Fig.~\ref{nuSK21}, we see that this picture is in very good agreement with 
the SK results. 

We are left with the constraints of KARMEN and LSND. 
The negative result of the KARMEN experiment can easily be accommodated as 
$x_{KARMEN}\simeq 3\cdot 10^{-4}\ x_{Chooz}$. 
On the contrary, as $x_{LSND}\simeq 6\cdot 10^{-4}\ x_{Chooz}$, our model can never 
comply with the LSND results, for any allowed values of $\theta $ and $\xi $.

We have thus shown that all experimental data (with the exception of LSND) 
can be accommodated in the simple 2-flavour model.
 
The fit could however be invalidated in the near future, should the LSND 
signal be confirmed by an independent experiment. A critical test will also 
be provided by the improving accuracy of the K2K experiment. Finally, the 2-flavour 
model could also be invalidated if $\nu _\tau$ are explicitly detected. We also point out 
that the astrophysical bound could be evaded in this model, since the disappearance 
of $\nu _e$ or $\nu _\mu$ in the extra dimensions is never complete (see~\cite{dienes}).

\section{Conclusions}

We have shown that, in the context of theories with large extra dimensions, 
a flavour neutrino coupled to a bulk fermion can manifest surprising and interesting 
behaviours, combining both oscillation and disappearance.
While a toy 1-flavour model is unable to fit all the experimental data, a simple
2-flavour model with only 3 free parameters meets most experimental constraints 
(except for LSND).

\section*{Acknowledgments}

Ling Fu-Sin benefits from a F.~N.~R.~S. grant.

\end{document}